\documentclass[aps,prl,amsmath,amssymb,floatfix, twocolumn]{revtex4-1}

\usepackage{multirow}
\usepackage{bbold}
\usepackage{graphicx}
\usepackage{subfigure}
\usepackage{color}
\usepackage{hyperref}
\usepackage[justification=raggedright]{caption}

\usepackage[normalem]{ulem}

\usepackage{physics}

\usepackage{amstext}
\usepackage{array}

\usepackage{diagbox}

\usepackage{siunitx}

\usepackage{yfonts}

\allowdisplaybreaks

\usepackage{setspace}

\newcommand{\ncmd}{\newcommand}
\ncmd{\nn}{\nonumber}
\ncmd{\pg}[1]{\textcolor{red}{#1}}
\ncmd{\mbf}[1]{\bs{#1}}
\ncmd{\Lam}{\Lambda}
\ncmd{\lam}{\lambda}
\ncmd{\Gam}{\Gamma}
\ncmd{\gam}{\gamma}
\ncmd{\sig}{\sigma}
\ncmd{\Dl}{\Delta}
\ncmd{\dl}{\delta}
\ncmd{\kap}{\kappa}
\ncmd{\mc}{\mathcal}
\ncmd{\veps}{\varepsilon}
\ncmd{\vphi}{\varphi}
\ncmd{\vtheta}{\vartheta}
\ncmd{\note}[1]{{\color{red}{\bf{#1} } } }
\ncmd{\new}[1]{{\texttt{#1}  } }
\ncmd{\eq}[1]{Eq. \eqref{#1}}
\ncmd{\bs}{\boldsymbol}
\ncmd{\pll}{\parallel}
\usepackage{natbib}
\usepackage{filecontents}

\begin{document}
\title{Topology of three-dimensional Dirac semimetals and generalized quantum spin Hall systems without gapless edge modes}
\author{Alexander C. Tyner$^{1}$} \thanks{These authors contributed equally.}
\author{ Shouvik Sur$^{2}$} \thanks{These authors contributed equally.}
\author{ Danilo Puggioni$^{3}$}
\author{James M. Rondinelli$^{1,3,4}$}
\author{Pallab Goswami$^{1,2}$}
\affiliation{$^{1}$ Graduate Program in Applied Physics, Northwestern University, Evanston, Illinois 60208}
\affiliation{$^{2}$ Department of Physics and Astronomy, Northwestern University, Evanston, Illinois 60208, USA}
\affiliation{$^{3}$ Department of Materials Science and Engineering, Northwestern University, Illinois 60208, USA}
\affiliation{$^{4}$ Northwestern Argonne Institute for Science and Engineering, Evanston, IL, 60208, USA}
\date{\today}

\begin{abstract}
Usually the quantum spin Hall states are expected to possess gapless, helical edge modes. Are there clean, non-interacting, quantum spin Hall states without gapless, edge modes? We show the generic, $n$-fold-symmetric, momentum planes of three-dimensional, stable Dirac semi-metals, which are orthogonal to the direction of nodal separation are examples of such generalized quantum spin Hall systems. We demonstrate that the planes lying between two Dirac points and the celebrated Bernevig-Zhang-Hughes model support identical quantized, non-Abelian Berry flux of magnitude $2 \pi$. Consequently, both systems exhibit spin-charge separation in response to electromagnetic, $\pi$-flux vortex. The Dirac points are identified as the unit-strength, monopoles of $SO(5)$ Berry connections, describing topological quantum phase transitions between generalized, quantum spin Hall and trivial insulators. Our work identifies precise bulk invariant and quantized response of Dirac semimetals and shows that many two-dimensional higher-order topological insulators can be understood as generalized quantum spin Hall systems, possessing gapped edge states.
\end{abstract}

\maketitle

\emph{Introduction}: The stable, three-dimensional, Dirac semimetals (DSM) arising from accidental linear touching between two Kramers-degenerate bands at isolated points in the Brillouin zone (BZ) are experimentally relevant examples of gapless topological states~\cite{wang2012dirac,wang2013three,yang2014classification,yang2015topological,Gorbar2015,Burkov2016,kargarian2016surface,gao2016classification,tang2016dirac,chiu2016classification,chang2017type,le2017three,le2018dirac,kargarian2018deformation,armitage2018weyl,kim2019,MaoLin2018,szabo2020,wieder2020}. The Dirac points of such systems, occurring along an $n$-fold axis of rotation are protected by the combined $\mc P \mc T$ and the $n$-fold, discrete, rotational ($\mathcal{C}_n$) symmetries, with $n= 3, 4, 6$~\cite{wang2012dirac,wang2013three}, where $\mc P$ and $\mc T$ represent space-inversion/parity ($\mathcal{P}$) and time-reversal ($\mathcal{T}$) symmetries, respectively. Several materials like Na$_3$Bi~\cite{wang2012dirac,liu2014discovery,xiong2015evidence,kushwaha2015bulk,liang2016electronic}, Cd$_3$As$_2$~\cite{wang2013three,liu2014stable,neupane2014observation,he2014quantum,borisenko2014experimental,moll2016transport,jeon2014landau}, PdTe$_2$~\cite{noh2017experimental}, $\beta'$-PtO$_2$~\cite{kim2019,wieder2020}, VAl$_3$~\cite{chang2017type}, $\beta$-CuI~\cite{le2018dirac}, KMgBi~\cite{wieder2020,le2017three}, PtBi$_2$~\cite{wu2019fragility}, and 
the magneto-electric (ME) compound FeSn~\cite{tang2016dirac,lin2020dirac} can host such Dirac points. Despite intensive theoretical research on stable DSMs for almost ten years~\cite{wang2012dirac,wang2013three,yang2014classification,yang2015topological,Gorbar2015,Burkov2016,kargarian2016surface,gao2016classification,tang2016dirac,chiu2016classification,chang2017type,le2017three,le2018dirac,kargarian2018deformation,armitage2018weyl,kim2019,szabo2020,wieder2020}, their bulk topological invariants are still unknown. 

The simplest version of DSMs can be obtained by stacking of Bernevig-Hughes-Zhang (BHZ) model~\cite{bernevig2006quantum} of quantum spin Hall (QSH) effect along the direction of nodal separation or the $\mathcal{C}_n$-axis. Since the BHZ model is a first order topological insulator (FOTI), supporting helical edge modes, the resulting DSM exhibits loci of zero-energy surface states, also known as the helical Fermi arcs. The total number of zero-modes is equal to the total QSH conductivity of DSMs, determined by $ \Delta k_D /\pi$, where $\Delta k_D$ is the distance between the bulk Dirac nodes. The spectroscopic and transport data of many stable DSMs are usually interpreted based on the existence of helical Fermi arcs ~\cite{liu2014discovery,xiong2015evidence,kushwaha2015bulk,liang2016electronic,liu2014stable,neupane2014observation,he2014quantum,borisenko2014experimental,moll2016transport,jeon2014landau,noh2017experimental,lin2020dirac}. 

However, recent theoretical works have showed that the generic, $n$-fold planes of DSMs are not described by the BHZ model possessing $U(1)$ spin-conservation law, or closely related $\mathbb{Z}_2$ FOTIs~\cite{Kane2005}. Away from the mirror planes, various crystalline-symmetry-preserving perturbations can gap out the helical edge modes. Using $K$-theory analysis, the generic planes were found to be topologically trivial~\cite{kargarian2016surface}. Subsequently, various groups~\cite{MaoLin2018, szabo2020, wieder2020} have identified these planes as higher-order, topological insulators (HOTI)~\cite{benalcazar2017}. The distinction between FOTI and HOTI is established by computing the nested Wilson loops of $SU(2)$ Berry connections for the occupied valence bands, under periodic boundary conditions. However, this difference only affects the physical properties under open boundary conditions, such as the presence of corner-states under $\mathcal{C}_n$-symmetric open boundary conditions. 

Are there any common topological properties shared by two-dimensional FOTI and HOTI under periodic boundary conditions? What happens to the QSH effect of the BHZ model, when the helical edge modes get gapped out by crystal-symmetry-preserving perturbations? We answer these two fundamental questions and identify the stable bulk topology of DSMs by performing second homotopy classification of non-Abelian Berry connections. 

\emph{Challenge toward topological classification}: The minimal model of a pair of two-fold, Kramers-degenerate bands of $\mc P \mc T$ symmetric systems is described by the Hamiltonian $H=\sum_{\bs{k}} \Psi^\dagger(\bs{k}) \hat{H}(\bs{k}) \Psi(\bs{k})$, where $\Psi(\bs{k})$ is a four-component spinor, and the Bloch Hamiltonian operator can be written as
$\hat{H}(\bs{k})= N_0(\bs k) \mathbb{1} + \sum_{j=1}^{5} \; N_j(\bs{k}) \Gamma_j$  ~\cite{Avron1988,Avron1989,demler1999,murakami2004}.
The magnitude of $O(5)$ vector field $\bs{N}(\bs{k})$ controls the spectral gap between conduction and valence bands, $N_0(\bs k)$ describes particle-hole anisotropy, and $\Gamma_j$ are five, mutually anti-commuting, $4 \times 4$ matrices, such that $\{\Gamma_i, \Gamma_j\}=2\delta_{ij}$. The topology of Bloch wave functions are determined by the unit, O(5) vector field $\hat{\bs{N}}(\bs{k})=\bs{N}(\bs{k})/|\bs{N}(\bs{k})|$, representing the coset space $SO(5)/SO(4)=S^4$, where $S^4$ is the unit, four-sphere. The diagonalizing matrix belongs to this coset space and the gauge group for intra-band Berry's connection is given by $Spin(4)=SU(2) \times SU(2)$. 

The vanishing of $|\bs{N}(\bs{k})|$ restores $SO(5)$-symmetry at the Dirac points, which serve as singularities of  $\hat{\bs{N}}(\bs{k})$. Whether the Dirac points are monopoles of Berry connection, leading to the quantized Berry flux for generic $n$-fold planes, can only be unambiguously determined by performing second homotopy classification of the gauge group. Since $\pi_2(S^4)$ and $\pi_2(SU(2) )\equiv \pi_2(S^3)$ are trivial, the homotopy analysis involves conceptual subtleties. We will show that the form of $\mc C_n$ operator can be exploited to identify a pair of global spin-quantization axes and reduce the redundancy of band eigenfunctions from $Spin(4)$ to $U(1) \times U(1)$, which allows second homotopy classification.

\begin{figure*}[!t]
\centering
\subfigure[]{
\includegraphics[width=0.3\textwidth]{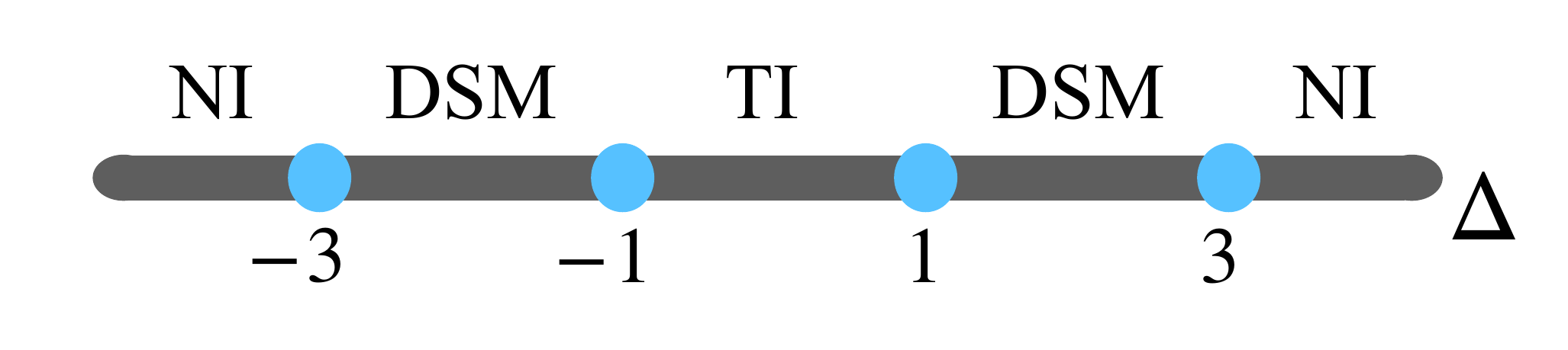}
\label{Fig1}}
\hfill
\subfigure[]{
\includegraphics[width=0.3\textwidth]{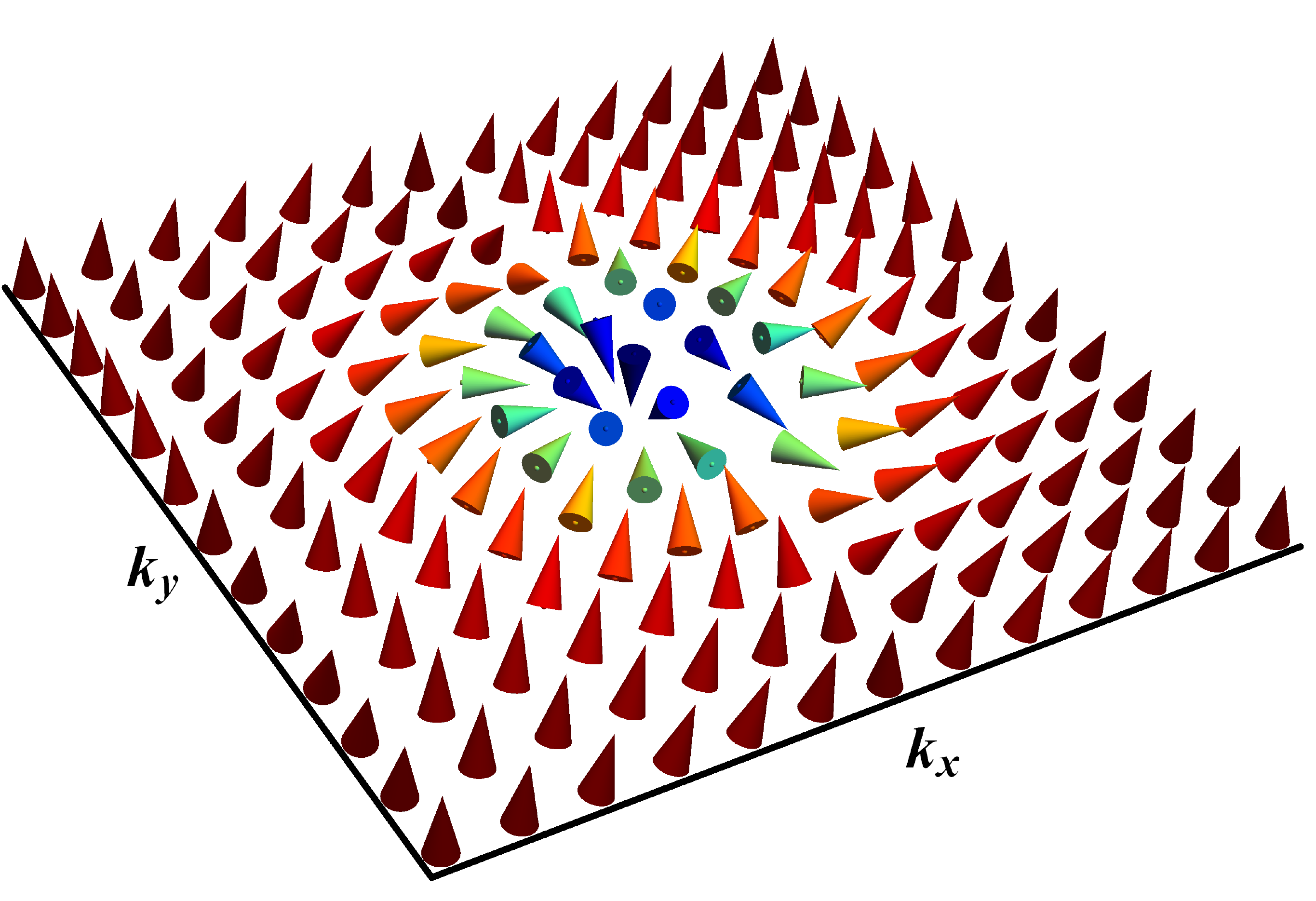}
\label{fig:3a}}
\hfill
\subfigure[]{
\includegraphics[width=0.3\textwidth]{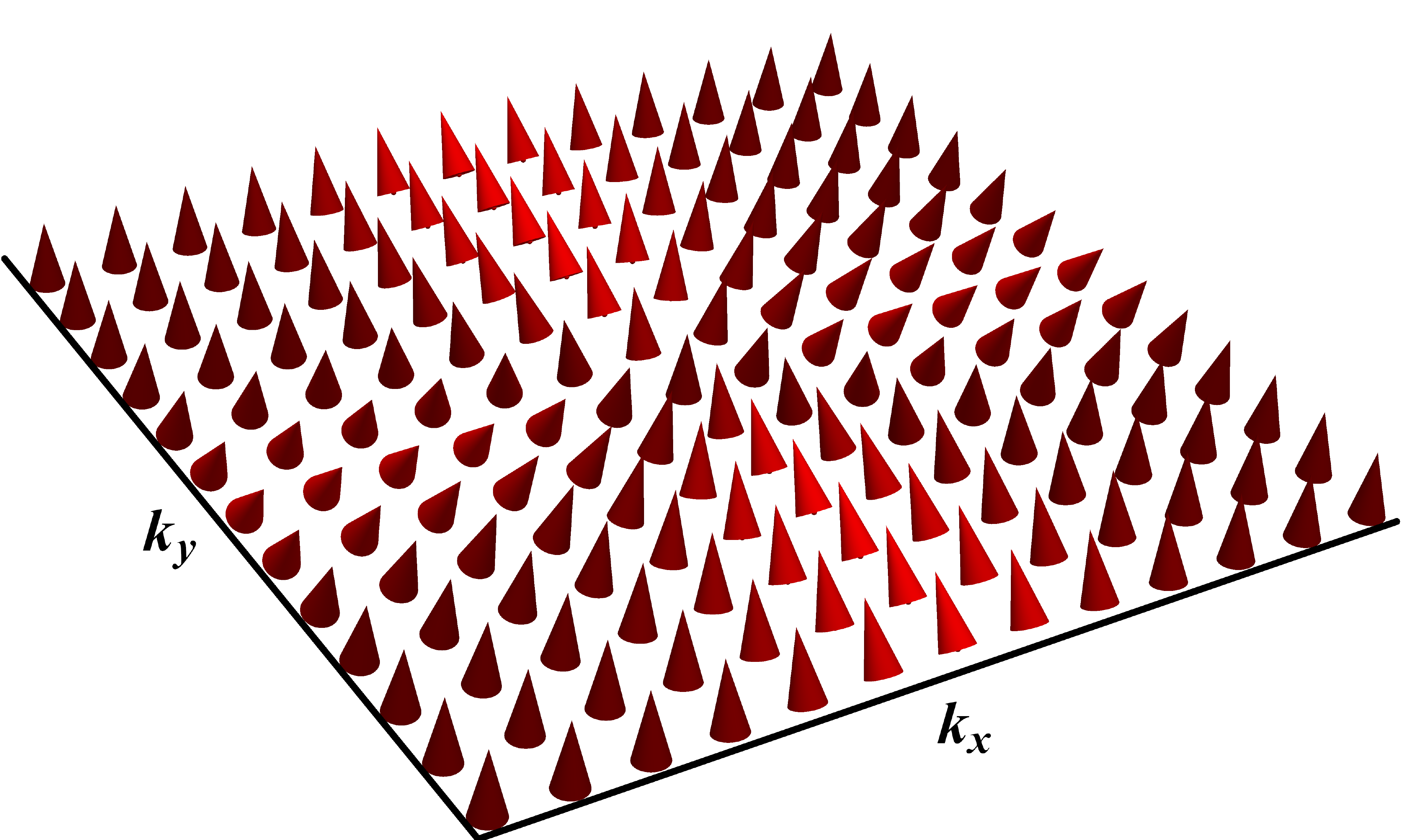}
\label{fig:3b}}
\hfill
\subfigure[]{
\includegraphics[width=0.3\textwidth]{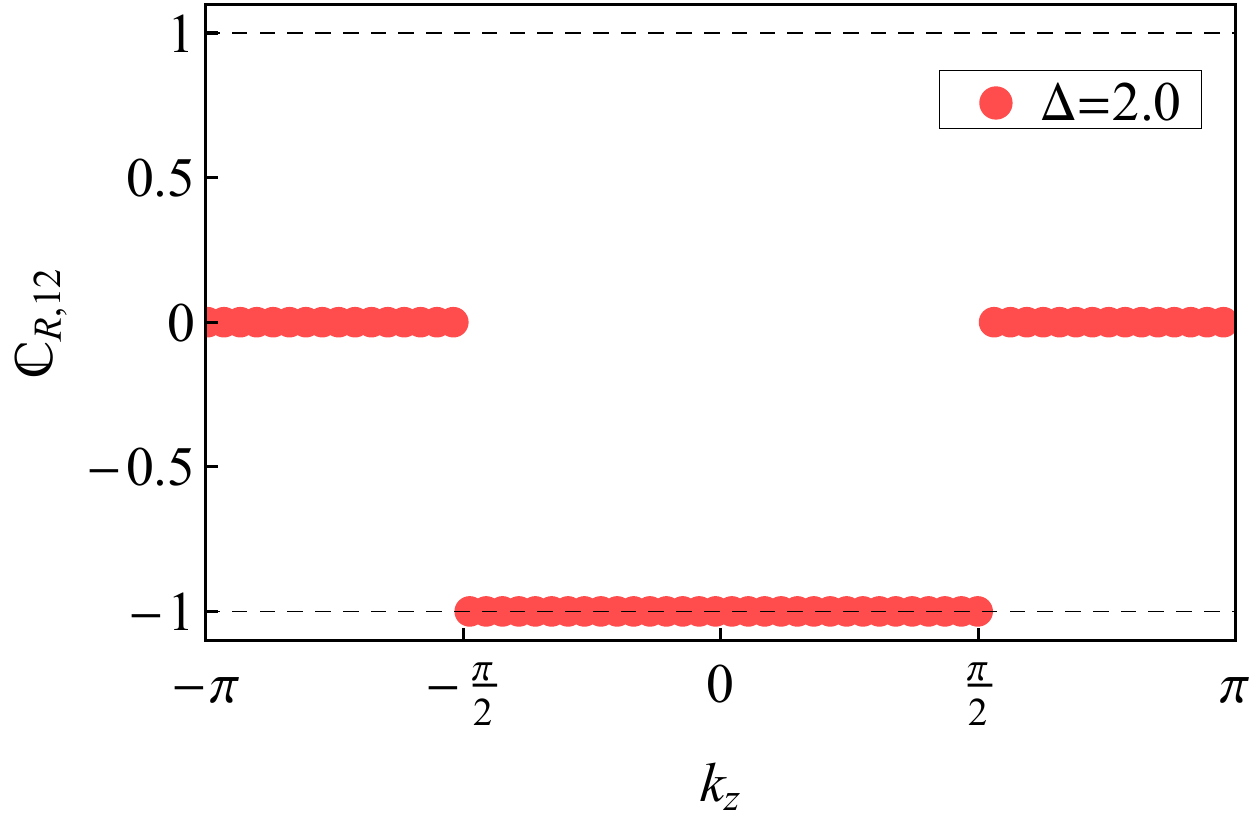}
\label{fig:3c}}
\hfill
\subfigure[]{
\includegraphics[width=0.27\textwidth]{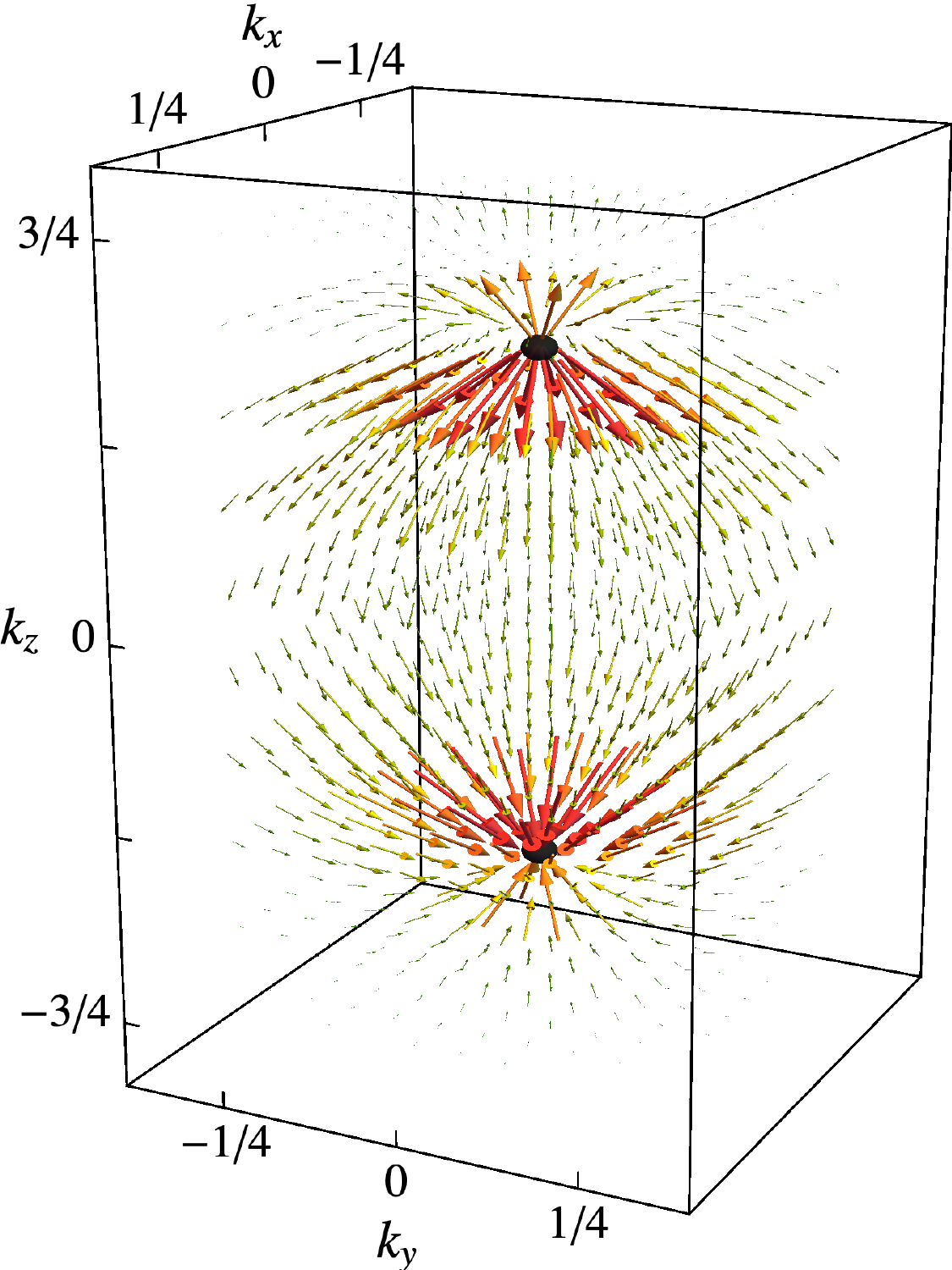}
\label{fig:3d}}
\hfill
\subfigure[]{
\includegraphics[width=0.3\textwidth]{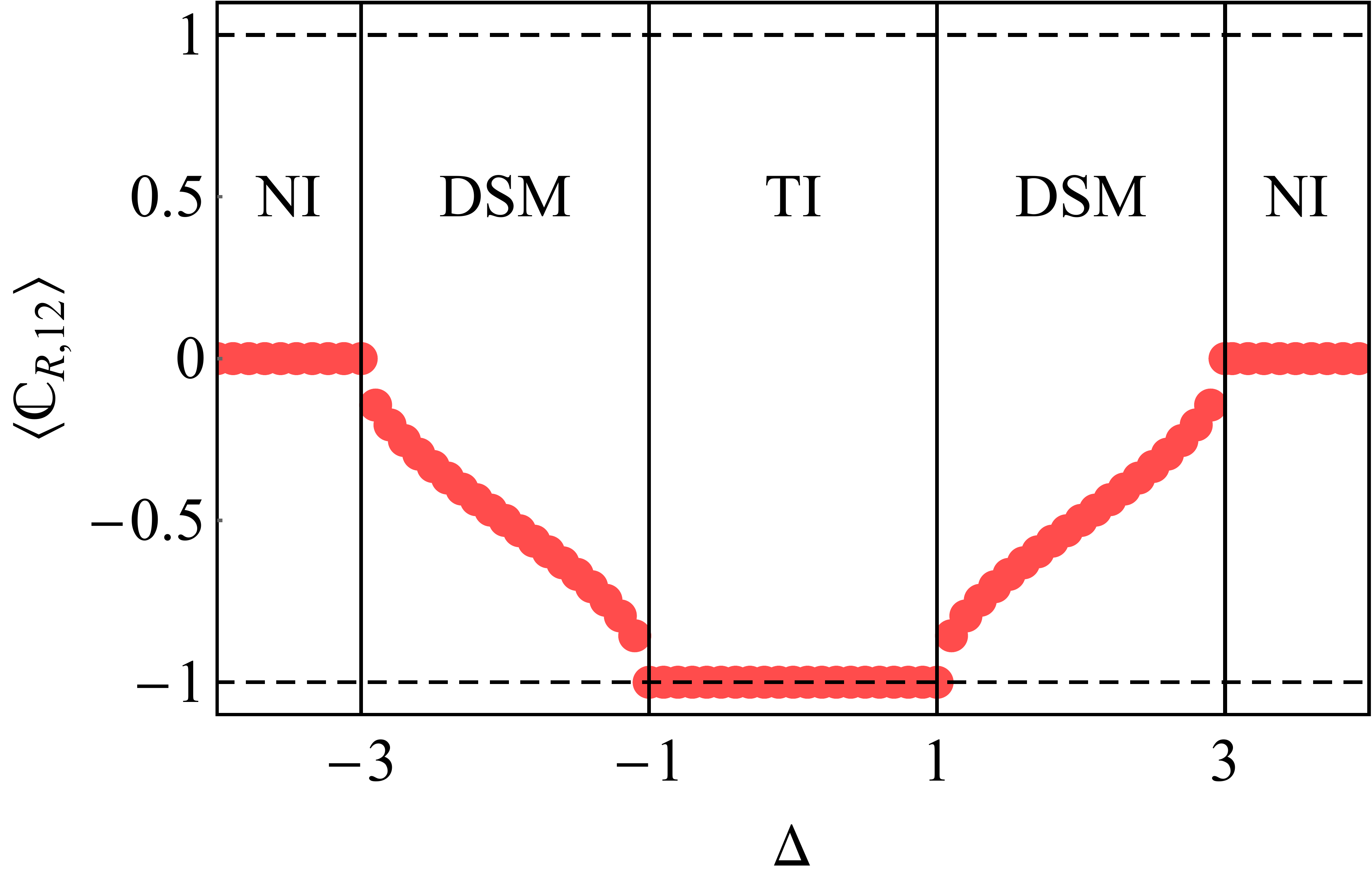}
\label{fig:3e}}
\caption{Phase diagram and bulk topology of DSM models. (a) Gapless topological phases (DSM), interpolate between a trivial/normal insulator (NI) and a topological insulator (TI), and 
topologically distinct phases are separated by quantum critical points (blue dots). 
(b) and (c): For all topologically non-trivial $4$-fold planes, the three-component, unit vector $\hat{\bs{n}}_{12}$($\hat{\bs{n}}_{34}$), defined by Eq.~(\ref{eq:UnitV}) displays skyrmion texture with winding number $-1(0)$.
(d) The relative Chern number or the quantized flux of $\bar{F}^{12}_{xy}(\bs{k})$ of Dirac semimetals for $\Delta=2$. 
(e) The vector plots of dipole configuration for Abelian projected magnetic fields $B^{12}_{i}(\bs{k})=\frac{1}{2} \epsilon_{ijl} \bar{F}^{12}_{jl}(\bs{k})$. The momentum components are in units of $\pi$. The Dirac points act as a pair of unit-strength, $SO(5)$ monopole and anti-monopole, where $\mathfrak C_{R,12} $ jumps by $\pm 1$. (f) The average value of the relative Chern number $\langle \mathfrak{C}_{R,12} \rangle$ per $xy$ plane, as a function of $\Delta$. }\label{Fig3}
\end{figure*}

\emph{Model}: We substantiate these claims by considering a model of $\mc C_4$-symmetric, magneto-electric DSMs, arising from the hybridization between $s$ and $p$ orbitals, \emph{which does not support any gapless surface states}. We will employ the following representation of gamma matrices $\Gamma_{j=1,2,3}=\tau_{1}\otimes \sigma_{j}$, $\Gamma_{4}=\tau_{2}\otimes \sigma_{0}$, $\Gamma_{5}=\tau_{3}\otimes \sigma_{0}$. The ten commutators $\Gamma_{jl}=[\Gamma_j,\Gamma_l]/(2i)$, with $j=1,...,5$ and $l=1,...,5$ serve as the generators of $SO(5)$ and its double cover group $Spin(5)$. The $2\times 2$ identity matrix $\tau_0$ ($\sigma_0$) and the Pauli matrices $\tau_j$'s ($\sigma_j$'s), with $j=1,2,3$ operate on orbital/parity (spin) index. The relevant $O(5)$ vector is given by
\begin{align}
&\bs{N}(\bs{k})=[t_p \sin k_x, t_p \sin k_y, t_{d,1}  (\cos k_x -  \cos k_y),  \nonumber \\ 
& t_{d,2} \sin k_x \sin k_y,  t_s(\Delta - \cos k_x - \cos k_y - \cos k_z)],
\label{O(5)}
\end{align}
where $t_s$, $t_p$, $t_{d,1}$, $t_{d,2}$ are four independent hopping parameters, and 
the dimensionless parameter $\Delta$ controls topological phase transitions. The phase diagram is shown in Fig.~\ref{Fig1}. The DSMs ($1<|\Delta | <3$) interpolate between trivial insulators ($|\Delta | >3$) and topological insulators ($|\Delta | <1$). We will focus on the parameter regime $1<\Delta<3$, with the Dirac points located at $\bs{k}_D=(0,0, \pm k_D)$, with $\cos(k_D)=(\Delta-2)$. Away from the high-symmetry locations $k_z=0$, $ \pi$, the generic $4$-fold planes of DSMs~\cite{yang2014classification}, preserving both $\mc P$ and $\mc T$ symmetries display identical form of $\bs{N}(\bs{k})$. 

\emph{Gauge-invariant Berry curvature}: The $\mc P \mc T$ symmetry is implemented by $\Gamma_{24} \hat{H}^\ast(\bs{k}) \Gamma_{24} = \hat{H}(\bs{k})$, and the diagonalizing matrix $U(\bs{k}) $ must satisfy the constraints $U^\dagger(\bs{k})  \hat{H}(\bs{k}) U(\bs{k}) = |\bs{N}(\bs{k})| \Gamma_5$, and $U^\dagger(\bs{k}) \Gamma_{24} U^\ast(\bs{k})= \Gamma_{24}$. Hence,  $U(\bs{k}) \in Spin(5)/Spin(4)$ has the general form~\cite{demler1999,murakami2004},  
\begin{eqnarray}
U(\bs{k}) &=& \begin{bmatrix}
\cos \frac{\theta(\bs{k})}{2} g_+(\bs{k}) & i \sin \frac{\theta(\bs{k})}{2} u(\bs{k}) g_-(\bs{k}) \\ 
i \sin \frac{\theta(\bs{k})}{2} u^\dagger(\bs{k}) g_+(\bs{k}) & \cos \frac{\theta(\bs{k})}{2} g_-(\bs{k})
\end{bmatrix} , \label{UG}
\end{eqnarray}
where the first (last) two columns correspond to the eigenfunctions of conduction (valence) bands. We have parametrized $S^4$ with a polar angle $\theta(\bs{k})$ and a four-component unit vector $\hat{\bs{n}}_\mu$, with $\mu=1,2,3,4$, such that $\cos[\theta(\bs{k})]= \frac{N_5(\bs{k})}{|\bs{N}(\bs{k})|}$, and $\hat{\bs{n}}_\mu=\frac{N_{\mu}(\bs{k})}{ |\bs{N}(\bs{k})|\sin[\theta(\bs{k})]}$. The $SU(2)$ matrix $u(\bs{k})=\hat{n}_4(\bs{k}) \sigma_0 + i \hat{n}_j(\bs{k}) \sigma_j$ describes the hybridization matrix elements between two orbitals, while two $SU(2)$ matrices $g_\pm(\bs{k})$ describe gauge freedom in selecting the eigenfunctions for conduction and valence bands, respectively. From $U(\bs{k})$ one finds the following intra-band $SU(2)$ connections 
\begin{eqnarray}
\bs{A}_+(\bs{k})&=&\sin^2 \frac{\theta}{2} \; g^\dagger_+ [- i u \nabla u^\dagger] g_{+} - i g^\dagger_+ \nabla g_+, \nonumber \\
\bs{A}_-(\bs{k})&=&\sin^2 \frac{\theta}{2} \; g^\dagger_- [- i u^\dagger \nabla u] g_{-} - i g^\dagger_- \nabla g_-,
\end{eqnarray}
for the conduction and valence bands, respectively. For notational compactness, we have suppressed the explicit $\bs{k}$-dependence of $\theta$, $u$, and $g_\pm$.

The $\mc C_4$ symmetry requires $\mc C_4 \hat{H}(\bs{k}) \mc C^\dagger_4 = \hat{H}(\bs{k}^\prime)$, which implements the constraint $[U^\dagger(\bs{k}^\prime) \mc C_4 U(\bs{k}), \Gamma_5 ] =0$, with the rotated wave vector $\bs{k}^\prime=(-k_y, k_x, k_z)$. For the orbital basis, $\mc C_4= e^{i \theta_p \sigma_3} \oplus e^{i \theta_q \sigma_3}$, with $\theta_p=\frac{\pi}{4}(2p+1)$, $\theta_q= \frac{\pi}{4}(2q+1) $, and $p=2 \; \mathrm{mod} \; 4$ and  $q=0 \; \mathrm{mod} \; 4$, and the hybridization matrix $u$ transforms as $u(\bs{k}^\prime)=e^{i \theta_p \sigma_3} u(\bs{k}) e^{-i \theta_q \sigma_3}$. In the band basis, the transformed rotation operator $\mc C_4^\prime (\bs{k}) \equiv  U^\dagger(\bs{k}^\prime) \mc C_4 U(\bs{k})$ becomes 
\begin{eqnarray}
\mc C_4^\prime (\bs{k}) = \qty[g^\dagger_+(\bs{k}^\prime) e^{i \theta_p \sigma_3} g_+(\bs{k})] \oplus \qty[g^\dagger_-(\bs{k}^\prime) e^{i \theta_q \sigma_3} g_-(\bs{k})], \nonumber \\
\end{eqnarray}
and the gauge choices $g_\pm (\bs{k})= \sigma_0$ and $g_\pm (\bs{k})= e^{i \alpha_\pm(\bs{k}) \sigma_3}$, keep the spin quantization axes unaffected.  

Any general choice of gauge specify a pair of local spin quantization axes $\hat{\bs m}_\pm (\bs{k})$, according to $g^\dagger_{\pm} (\bs{k})\sigma_3 g_{\pm} (\bs{k})= \hat{\bs m}_\pm (\bs{k}) \cdot \boldsymbol \sigma$. Once $g_\pm(\bs{k}) $ are identified, $U(\bs{k})$ only exhibits residual $U(1) \times U(1)$ gauge freedom, corresponding to the spin rotations about $\hat{\bs m}_\pm (\bs{k})$, i.e.,  $g_{\pm} (\bs{k}) \to g_\pm(\bs{k}) \exp\qty[i \varphi_{\pm} (\bs{k} ) \hat{\bs m}_\pm (\bs{k}) \cdot \boldsymbol \sigma ]$. Consequently, the gauge group of intra-band Berry connection is given by $Spin(4)/[U(1) \times U(1)]$, with the second homotopy class 
\begin{align}
\pi_2 \left(\frac{Spin(4)}{U(1) \times U(1)} \right)=\pi_1(U(1) \times U(1))=\mathbb{Z} \times \mathbb{Z}. \label{equivalence}
\end{align} 
Hence, the topology of $n$-fold planes and the Dirac points are governed by a pair of integer invariants, and \emph{the Dirac points can be identified as non-Abelian monopoles}. 

 The Abelian projected Berry connections can be obtained as $\bar{\bs{A}}_\pm (\bs{k})= \frac{1}{2} Tr[\bs{A}_\pm (\bs{k})\hat{\bs m}_\pm (\bs{k}) \cdot \boldsymbol \sigma]=\frac{1}{2} Tr[\bs{A}_\pm (\bs{k})g^\dagger_{\pm} (\bs{k})\sigma_3 g_{\pm} (\bs{k})]$, leading to
\begin{eqnarray}
\bar{\bs{A}}_+ (\bs{k})=\frac{1}{2} \sin^2 \frac{\theta}{2} \; Tr[ - i u \nabla u^\dagger \sigma_3] + \frac{i}{2} Tr[g^\dagger_+ \nabla g_+ \sigma_3], \nonumber \\
\bar{\bs{A}}_- (\bs{k})=\frac{1}{2} \sin^2 \frac{\theta}{2} \; Tr[ - i u^\dagger \nabla u \sigma_3] +\frac{i}{2} Tr[g^\dagger_-\ \nabla g_- \sigma_3]. \nonumber \\
\end{eqnarray}
Consequently, the gauge-invariant, quantized Berry flux can be determined from the Abelian field strength tensors (or Berry curvatures) $\bar{F}_{ij,\pm}(\bs{k}) = \partial_i \bar{A}_{j,\pm}(\bs{k})-\partial_j \bar{A}_{i,\pm}(\bs{k})$. For all smooth gauge transformations, such that the spin quantization axes are topologically trivial, meaning the gauge-fixing operators $\hat{\bs m}_\pm (\bs{k}) \cdot \boldsymbol \sigma$ do not correspond to fictitious two-band models of Chern insulators, $i/2 Tr[g^\dagger_\pm \nabla g_\pm \sigma_3]$ terms cannot contribute to the quantized flux of $\bar{F}_{ij,\pm}(\bs{k})$ or the relative Chern numbers for $4$-fold planes, defined as
\begin{align}
\mathfrak C_{R, \pm}(k_z) = \frac{1}{2\pi} \; \int_{T^2} \; dk_x dk_y \; \bar{F}_{xy,\pm}(\bs{k}).
\end{align}

\emph{Quantized Berry flux}:  Next, we perform explicit analytical calculations of Berry flux with the global gauge choice $g_\pm (\bs{k})= \sigma_0$, corresponding to the spin quantization axes $\hat{\bs m}_\pm (\bs{k})=(0,0,1)$. 
It is convenient to define symmetric and anti-symmetric combinations of Berry curvatures as $\bar F_{\ij}^{12}=(\bar F_{\ij, +}+\bar F_{\ij, -})/2$, and $\bar F_{\ij}^{34}=(\bar F_{\ij, +}-\bar F_{\ij, -})/2$. These curvatures will be associated with the diagonal, Cartan generators of $SO(5)$ group, namely $\Gamma_{12}=\tau_0 \otimes \sigma_3$ and $\Gamma_{34}=\tau_3 \otimes \sigma_3$, and can be elegantly written as $ \bar{F}^{ab}_{ij} = \sin(\theta_{ab})[\partial_i \theta_{ab} \partial_j\phi_{ab}-\partial_j \theta_{ab} \partial_i\phi_{ab}]$, 
where we have introduced two sets of spherical polar angles $(\theta_{12}(\bs{k}), \phi_{12}(\bs{k}))$ and  $(\theta_{34}(\bs{k}), \phi_{34}(\bs{k}))$, such that 
\begin{eqnarray}
&&\tan[\phi_{ab}(\bs{k})]=\frac{N_b(\bs{k})}{N_a(\bs{k})}, \\ && \cos[\theta_{ab}(\bs{k})]=1-\frac{[N^2_a(\bs{k}) + N^2_b(\bs{k})]}{|\bs{N}(\bs{k})|[|\bs{N}(\bs{k})|+N_5(\bs{k})]}.
\end{eqnarray}
The quantized flux of $\bar{F}^{12}_{ij}$ and $\bar{F}^{34}_{ij}$ can only exist if BZ two-torus can wrap around unit two spheres, defined by
\begin{equation} \hat{\mathbf{n}}_{ab}=(\sin \theta_{ab} \cos \phi_{ab},  \sin \theta_{ab} \sin \phi_{ab} , \cos  \theta_{ab} ).\label{eq:UnitV}\end{equation}
Notice that $\Phi^{12}_{xy}(k_z)=2\pi \mathfrak C_{R,12}(k_z) $ and $\Phi^{34}_{xy}(k_z)=2\pi \mathfrak C_{R,34}(k_z) $ describe the flux of Abelian fields $\bar{F}^{12}$ and $\bar{F}^{34}$, respectively.

\begin{figure}[!t]
\centering
\subfigure[]{
\includegraphics[scale=0.5]{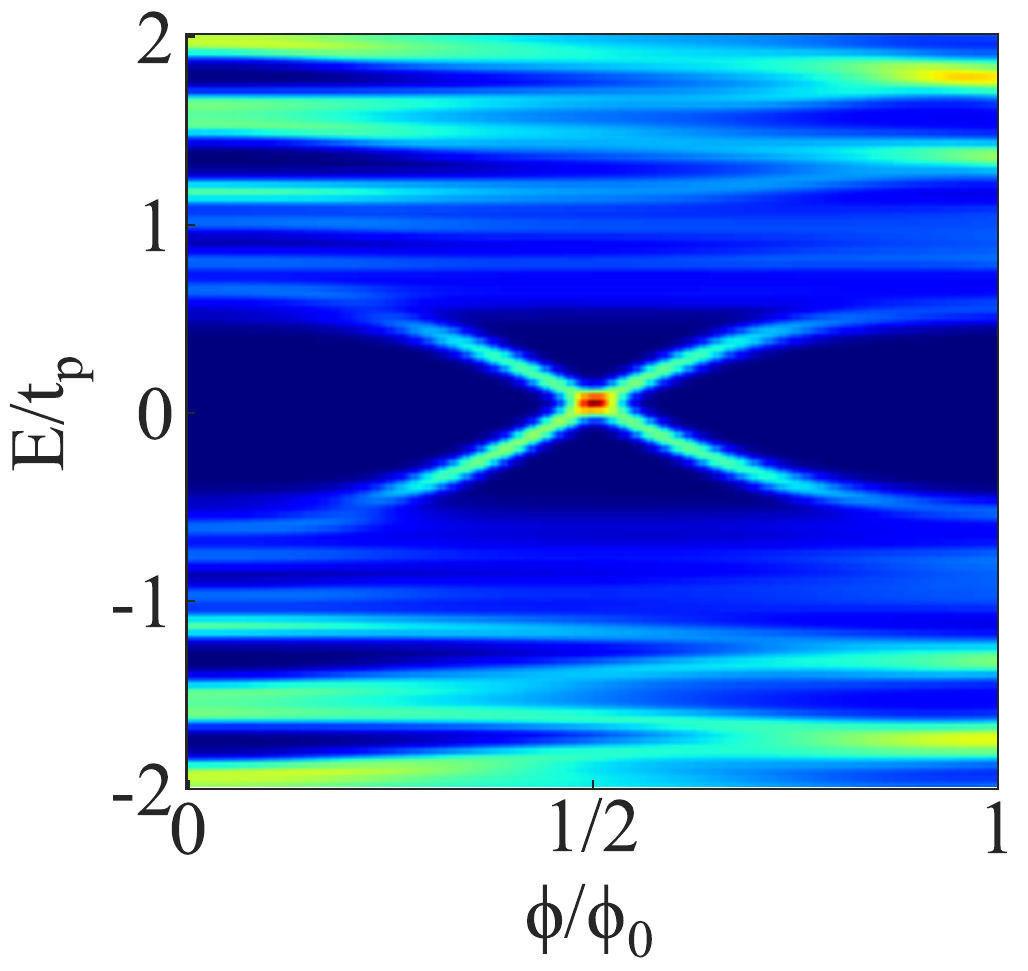}
\label{fig:Flux_States}}
\hfill
\subfigure[]{
\includegraphics[scale=0.45]{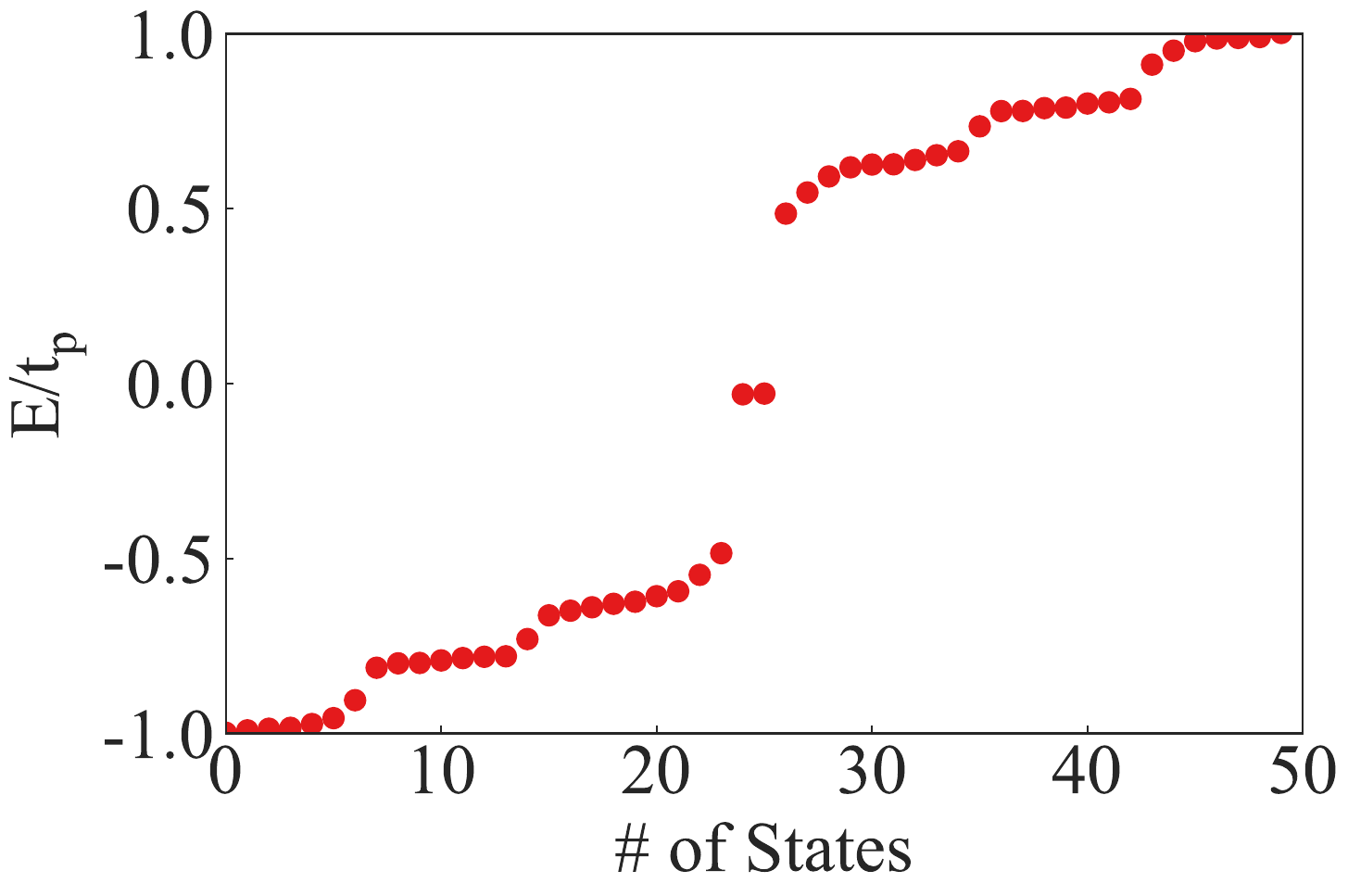}
\label{fig:BHZFlux}}
\caption{Non-perturbative signatures of quantum spin Hall effect for topologically non-trivial planes of Dirac semimetals. (a) Local density of states at the location of an electromagnetic flux tube, as a function of energy $E$ (measured in units of hopping parameter $t_p$) and the strength of flux $\phi$, and $\phi_0=hc/e$ is the flux quantum. (b) The number of states vs. energy eigenvalues for $\phi=\phi_0/2$, showing the higher-order topological insulators, described by the generic planes of Dirac semimetals support two-fold degenerate, zero-energy states. }\label{fig:Flux_Insertion}
\end{figure}

For all topologically non-trivial $4$-fold planes of $\mc C_4$-symmetric DSMs described by Eq.~(\ref{O(5)}), only $\theta_{12}$ interpolates between $0$ and $\pi$, leading to the skyrmion configuration for the unit vector 
$\hat{\bs{n}}_{12}$, as shown in Fig.~\ref{fig:3a}. 
In contrast to this, $\theta_{34}$ does not interpolate between $0$ and $\pi$, and the corresponding unit vector $\hat{\bs{n}}_{34}$ is topologically trivial, as shown in Fig.~\ref{fig:3b}. 
The quantization of the relative Chern numbers, and their discontinuities at the Dirac points are shown in Fig.~\ref{fig:3c}. 
The monopole numbers for the Dirac points at $\bs{k}=(0,0, \pm k_{D,j})$ are determined by $\mc N_{12}(\pm k_{D,j}) = 
\lim_{\epsilon \to 0} [\mathfrak C_{R,12}(k_z=  \pm k_{D,j}+\epsilon) -\mathfrak C_{R,12}(k_z=  \pm k_{D,j}-\epsilon)] = \pm 1$, and  
$\mc N_{34}(\pm k_{D,j}) = 0$.
In Fig.~\ref{fig:3d}, we illustrate the structure of Abelian projected magnetic fields $B^{12}_{i}(\bs{k})=\frac{1}{2} \epsilon_{ijl} \bar{F}^{12}_{jl}(\bs{k})$, which support dipole configuration.
Using the $k_z$-dependent relative Chern numbers, we can also define the average relative Chern numbers per $xy$ plane $\langle \mathfrak C_{R,ab} \rangle (\Delta)  = \frac{1}{2\pi} \; \int_{-\pi}^{\pi} dk_z \; \mathfrak C_{R,ab}(k_z)$, 
which is shown in Fig.~\ref{fig:3e}. We note that the stacked BHZ model with $t_{d,1}=t_{d,2}=0$, the stacked HOTI with $t_{d,2}=0$\cite{benalcazar2017}, and the stacked HOTI with $t_{d,1}=0$ support identical quantized flux of $\bar{F}^{12}_{jl}(\bs{k})$. 
Hence, the relative Chern number acts as a topological order parameter for various phases, controlling the strength of generalized QSH effect, which can be seen in the following manner. 

\emph{Generalized QSH effect}: Refs.~\onlinecite{{QiSpinCharge,RanSolitons,JuriProbes,mesaros2013zero}} have identified spin-charge separation as the non-perturbative signature of QSH, which can survive as a genuine topological response even in the absence of $U(1)$ spin conservation law. For the BHZ model ($t_{d,1/2}=0$) and closely related $Z_2$ FOTI, supporting gapless edge modes, it was shown that an electromagnetic $\pi$ flux tube binds two-fold degenerate, zero-energy, mid-gap states. When both states are occupied (empty), the Kramers-singlet, ground state carries charge +e (-e). In contrast to this, the half-filling of zero-modes corresponds to Kramers-doublet with charge $e=0$. 

To demonstrate spin-charge separation for $\mathcal{C}_4$-symmetric HOTI, we have computed the local density of states in the presence of an electromagnetic flux tube, oriented along the $z$-axis, for a system size $21 \times 21$, under periodic boundary condition. The local density of states at the location of flux tube is shown in Fig.~\ref{fig:Flux_States} as a function of energy and the strength of flux $\phi$. The calculations were performed with hopping parameters $t_s=t_p=t_{d,1}=t_{d,2}$, $k_z=\pi/2$, and $\Delta=1.5$. The low-energy states for $\phi=\phi_0/2$ i.e., $\pi$-flux are shown in Fig.~\ref{fig:BHZFlux}, providing clear evidence for the existence of two-fold degenerate, mid gap states at zero-energy. All topologically non-trivial planes of DSMs can support such mid-gap states (which may or may not be at zero energy), and their total number corresponds to $ \Delta k_D /\pi$.  \emph{Therefore, the relative Chern number provides a unified theoretical framework for describing generalized QSH effect of Kramers-degenerate FOTI and HOTI, irrespective of the presence or absence of gapless edge-modes and corner-localized states}. 

\acknowledgements{
This work was supported by the National Science Foundation MRSEC program (DMR-1720139) at the Materials Research Center of Northwestern University, and the start up funds of P. G. provided by the Northwestern University. P. G. completed a part of this work at the Aspen Center For Physics, which is supported by National Science Foundation grant PHY-1607611. D.P. and J.M.R. acknowledge the Army Research Office under Grant No. W911NF-15-1-0017 for  financial support and the DOD-HPCMP for computational resources.
}

\bibliography{Ref.bib}

\end{document}